# Integrated Multiphysics Modeling of a Piezoelectric Micropump

**AmirHossein Ghaemi[1], Abbas Ebrahimi[2], Majid Hajipour[3]**
1- M.Sc. Student, Faculty of Aerospace Engineering, Delft University of Technology, Delft, Netherlands
2- Associate Professor, Aerospace Engineering Department, Sharif University of Technology, Tehran, Iran
3- Assistant Professor, College of Interdisciplinary Science and Technology, University of Tehran, Tehran, Iran

**Abstract**
This paper presents an integrated multiphysics simulation approach of piezoelectric micropumps. Micropumps and micro blowers are essential devices in various cutting-edge industries like laboratory equipment, medical devices, and fuel cells. A piezoelectric micropump involves complex physics including microfluidics, flow-structure interaction, electricity, and piezoelectric material. Hence, a comprehensive analysis of the interactions between different physical phenomena, would be essential for the effective design and optimization of these micropumps. Prior studies on piezoelectric micropump were mainly focused on isolated physical aspects of these pumps, such as piezoelectric mechanics, fluid dynamics, electrical properties, and also fluid-structure Interactions. The present paper fills this gap by integrating these aspects into a holistic simulation and design approach, introducing a new methodology for micropump analysis. Advanced simulation and design tools like COMSOL and SolidWorks were employed in accordance. A brief review of piezoelectric materials, and an exploration of different types of micropumps and their operating principles is discussed. Also, a comparison of various piezoelectric materials, including their properties and applications is investigated. Further, the paper discusses the simulation process of the micropumps, using COMSOL software, and presents an in-depth analysis of the simulation results. This structured approach provides a comprehensive understanding of piezoelectric micropumps, from theoretical underpinnings to practical design considerations.

## 1. Introduction

Micropumps and micro blowers are essential devices in various cutting-edge industries like laboratory equipment, medical devices, and fuel cells. A piezoelectric micropump involves complex physics including microfluidics, flow-structure interaction, electricity, and piezoelectric material. Hence, a comprehensive analysis of the interactions between different physical phenomena, would be essential for the effective design and optimization of these micropumps. Ulmann [1], Investigated the performance of single and dual-chamber valveless piezoelectric pumps. It was found that a series arrangement of these pumps, despite equal pressure differences, yielded a higher mass flow rate compared to a parallel arrangement. Li, et al. [2] Examined high-flow, stable, passive microvalves for piezoelectric pump-based actuators. These valves utilized flaps to enhance performance up to 10 kilohertz frequency and 10 MPa pressure.

Wu, et al. [3] focused on PZT micro pumps with low power consumption, particularly for fluid transport in Direct Methanol Fuel Cells systems. The studied pump included two inlet and outlet valves, a chamber, a diaphragm, and an actuator. Cheng, et al. [4] analyzed a piezoelectric disc using the finite element method. This pump's inlet and outlet valves were opened by magnetic coils and closed by fans.

Kim, et al. [5] designed a valveless piezoelectric pump for micro-fluidic devices. This numerical simulation involved a pump with two coaxial cylindrical shells connected to a piezoelectric ceramic ring and a metal body, using a rotating wave in an ultrasonic motor. De Lima, et al. [6] developed a pump using biomorphic piezoelectric plates, inspired by fish swimming mechanics. This study, both numerical and experimental, utilized body and tail fin movements to generate fluid flow.

Eastman, et al. [7] experimentally investigated the cooling potential of a piezoelectric air pump on a heating plate. Various parameters like the pump's distance from the heating plate, plate diameter, and air blowing time were examined. The results indicated that the heat transfer coefficient is dependent on distance and voltage applied to the piezoelectric material. Cazorla, et al. [8] developed and analyzed a micro pump using silicon and PZT piezoelectric plates, suitable for both water and air. One of its advantages was low power consumption.

Chen, et al. [9] examined a piezoelectric pump for medical applications, specifically drug delivery. This theoretical and practical study revealed that diaphragm thickness significantly affects the pump's optimal frequency, and the chamber height influences the maximum flow rate.

Prior studies on piezoelectric micropump were mainly focused on isolated physical aspects of these pumps, such as piezoelectric mechanics, fluid dynamics, electrical properties, and also fluid-structure Interactions. The present paper fills this gap by integrating these aspects into a holistic simulation and design approach, introducing a new methodology for micropump analysis.

This paper focuses on Multiphysics simulation of piezoelectric micropumps. It emphasizes the integration of diverse physical phenomena—like piezoelectric mechanics and fluid dynamics—using simulation tools

2. ebrahimi_a@sharif.ir (corresponding author)

to enhance the design and efficiency of these micropumps. This approach represents a stride in micropump technology, offering new insights and methodologies for their development.

**2. Piezoelectric Micropump Overview**
In this section, preliminary studies related to the paper are presented, along with a brief overview of various types of piezo electrics. Subsequently, the chapter introduces different types of micropumps and provides a concise description of each. Following this, the history of piezo pumps is examined. Finally, a piezoelectric air pump is selected for simulation purposes.

**1) Review of piezoelectric materials:** piezo electrics, a category of smart materials, uniquely convert mechanical energy to electrical energy and vice versa. This property was first identified in specific crystals and later mathematically deduced for the reverse effect.

The direct piezoelectric effect involves electric polarization that is proportional to mechanical strain, thereby generating electricity in the material under stress. Conversely, the reverse piezoelectric effect occurs when an applied voltage induces mechanical strain proportional to the electric field's magnitude. Piezoelectric materials are classified into five types:

1) Piezoelectric Crystals: These include materials like quartz and are used in various applications such as watches, radios, and microphones. They are known for their durability but have relatively low sensitivity.

2) Piezoelectric Semiconductors: This category includes certain semiconductor materials that have been instrumental in advancing electro-acoustic devices.

3) Piezoelectric Ceramics: Discovered post-World War II, these multi-crystalline materials like $BaTiO_3$ and PZT are cost-effective and easily shaped. They find applications in sound wave generation and various sensor technologies.

4) Piezoelectric Polymers: In response to the need for larger and more flexible materials, polymers were studied and found to exhibit desired piezoelectric properties. They are used in medical equipment, robotics, and other fields requiring flexibility and lightweight materials.

5) Piezoelectric Composites: These are combinations of piezoelectric ceramics with non-piezoelectric materials, aimed at improving mechanical properties for specific applications. They are utilized in ultrasonic imaging and medical devices.

**2) Review of piezoelectric micropumps:** In this subsection, various mechanisms used for pumping fluids at a micro-scale are discussed. Micro pumps are classified based on the type of driving force and the nature of fluid inlet and outlet. Key performance parameters and general characteristics of micro pumps are presented.

Micro pumps typically use positive displacement via a cavity or chamber. Volume changes in two phases: Supply (increasing volume) and Pump (decreasing volume) phases. Net flow rate and head pressure are crucial performance parameters. The efficiency of pumps is defined by the maximum back pressure they can withstand while maintaining flow and the relationship between back pressure and flow rate is inverse. Classification Based on Driving Force:

1) Electrostatic Force: This type utilizes oppositely charged electrodes. Advantages include ease of operation and cost-effectiveness; drawbacks are high required DC voltage and potential clogging.

2) Piezoelectric Effect: This type involves piezoelectric actuators, typically made of PZT ceramic. Benefits include low power consumption and precise control; challenges are high voltage requirement and limited actuator displacement.

3) Pneumatic Force: This type uses gases that expand significantly upon heating. The main advantage is high displacement; drawbacks include low frequency and the possibility of bubble formation.

4) Variable Electromagnetic Field: This type employs alternating current in micro coils. Advantages are good displacement and force; limitations include the placement of coils and magnets, restricting size.

Classification Based on Inlet/Outlet Flow:

1) Valveless Micro Pumps: This type use diffuser and nozzle mechanisms. Efficiency is low due to high backflow, but their construction is simpler.

2) One-Way Valve or Passive Micro Pumps: This type utilizes one-way valves like flap-like valves. Challenges include construction complexity and self-priming issues.

3) Active Valve Micro Pumps: This type is Complex in design and construction, not commonly used due to high cost and control difficulties.

**3. Integrated Multiphysics Simulation**
This section provides an overview of the development process of the piezoelectric micropump, encompassing design principles, simulation techniques, and material selection. It emphasizes the role of design in optimizing pump efficiency and functionality.

**3.1 Computer Aided Design**
In the computer-aided design section, it's noted that from an original reference design, the concept was simplified into two distinct models for Multiphysics simulation purposes. The first model is a 3D axisymmetric design, which offers a more comprehensive representation.

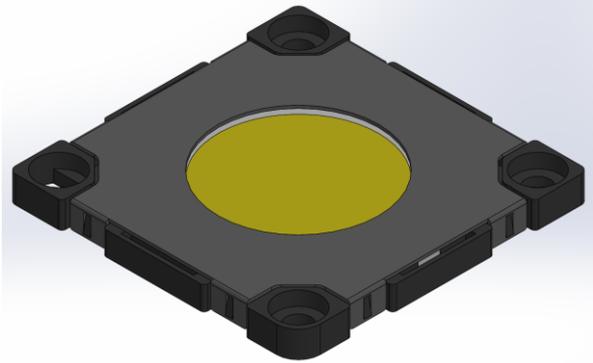

Fig. 1: CAD of Original Reference Design

The second is a 2D model, chosen primarily to reduce computational costs while still providing valuable insights into the system's behavior. These adaptations allow for efficient exploration of the design space while managing the resources required for simulation.

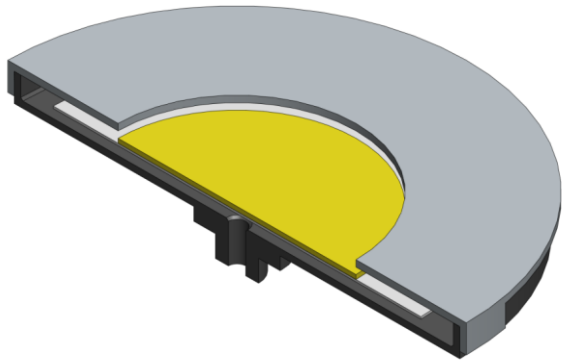

Fig. 2: Section View of 3D Axisymmetric Design CAD

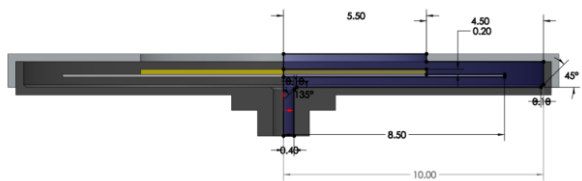

Fig. 3: Obtaining 2D Design from Fig. 2

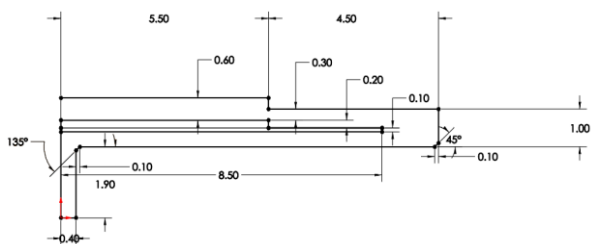

Fig. 4: Final 2D Design

In the development of a new piezoelectric pump, the design initially takes inspiration from Murata company's existing pump models. The design strategy opts for an axisymmetric approach, simplified for ease of analysis and prototype fabrication, with zero displacement constraints assumed for the surrounding steel structure.

This leads to a consistent axisymmetric motion, facilitating efficient design and analysis processes.

The piezoelectric component's thickness is a critical factor, determined with reference to industrial standards. While the ideal thickness for maximum displacement is around 0.1 mm, practical limitations necessitated the use of a slightly thicker material at 0.2 mm. The diameter of the piezoelectric component mirrors that of the reference model at 11 mm. Similarly, the metallic component is designed with a diameter of 17 mm and a thickness of 0.1 mm.

### 3.2 Modal Analysis

In the initial phase of our study, a comprehensive modal analysis was performed using COMSOL Multiphysics software. This analysis plays a pivotal role in assessing the dynamic characteristics of the piezoelectric pump. For this purpose, a finely tuned mesh was developed. The mesh settings were adjusted to 'Physics Controlled' and 'Fine' to ensure the accuracy and reliability of the analysis.

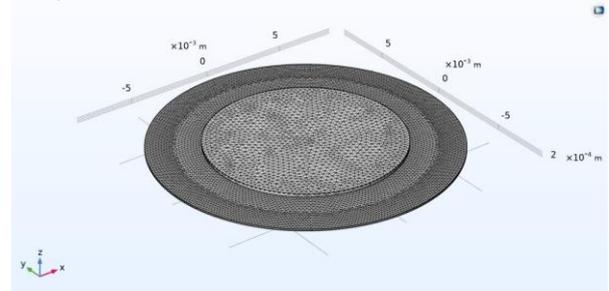

Fig. 5: Modal Analysis Mesh

The core objective of this phase was to identify the first 20 modal frequencies, with a particular focus on frequencies in the vicinity of 1 kHz. A key finding from this analysis was the identification of a peak displacement at the center of the pump at a frequency of 4.288 kHz. This specific frequency emerged as a critical parameter and was subsequently chosen for further analysis. This choice is instrumental in guiding the design process, ensuring the optimization of the pump's vibrational efficiency and overall performance.

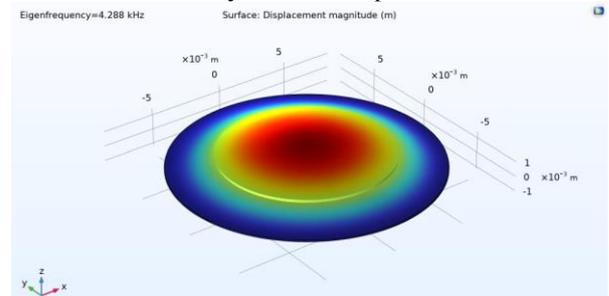

Fig. 6: Selected Mode Shape

### 3.3 Time Dependent Structural Modeling

The study progresses with a time-dependent analysis. This analysis was first carried out on a standard computer for a shorter duration and subsequently executed on a High-Performance Computing (HPC) system for a longer period. The initial solution was conducted with time step of one-twentieth of an oscillation period, and the solution was computed for 50 oscillation periods. The maximum time step was kept constant and equal to the specified time step. The initial time step was set to one-twentieth of an oscillation period as well. In the relevant section, the 'Maximum step constraint' was altered to a constant state.

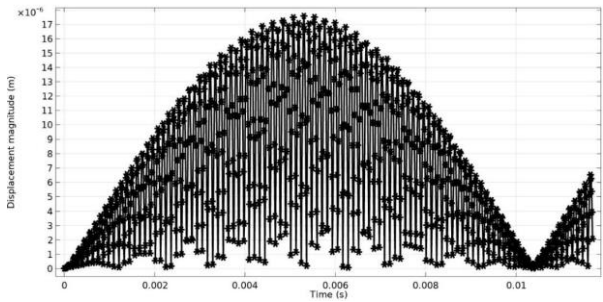

Fig. 7: Maximum Displacement for the Upper Surface (Second Run)

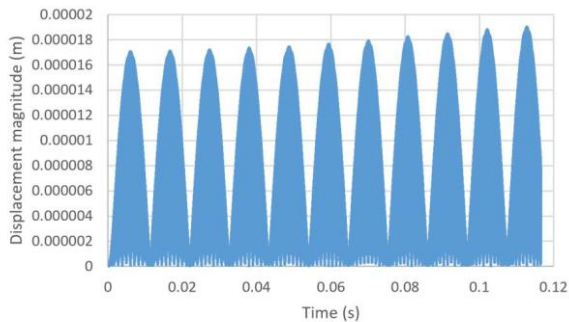

Fig. 8: Maximum Displacement for the Upper Surface (Second Run)

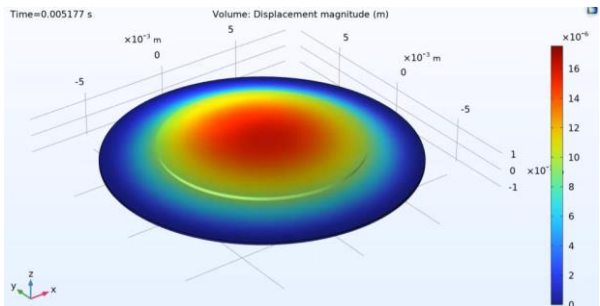

Fig. 9: Frame of Piezoelectric Movement

It was observed that the phenomenon of Beating occurred in the system. The Beating frequency is significantly lower than the oscillation frequency, yet it has a considerable impact on the amplitude of the piezo pump's motion. Due to the absence of damping, the system's amplitude does not stabilize, resulting in the beating condition. Due to the significance of damping in the dynamic behavior of the system [10], damping properties were introduced to both the steel and piezoelectric. The type of damping was set to Rayleigh damping. It was identified that for each of the two natural frequencies of the system, corresponding damping coefficients needed to be inputted. By referencing the analysis of the model and obtaining the value of the next natural frequency, both the oscillation frequency and this newly identified frequency were inputted as the required natural frequencies. Assuming that the damping coefficients are the same at these two frequencies, values of 0.01, 0.03, and 0.05 were sequentially inputted and tested. The results of these tests are as follows:

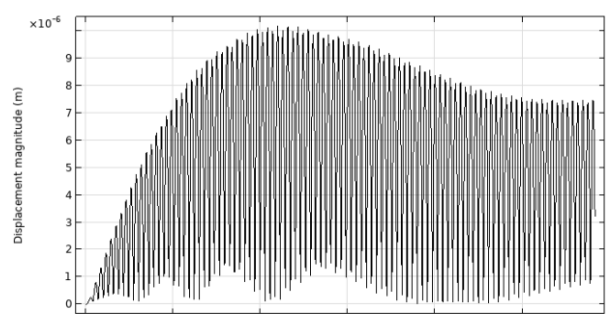

Fig. 10: Maximum Displacement for the Upper Surface (Damping Coefficient = 0.01)

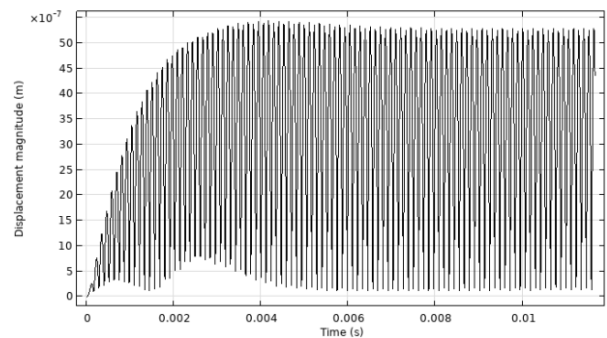

Fig. 11: Maximum Displacement for the Upper Surface (Damping Coefficient = 0.03)

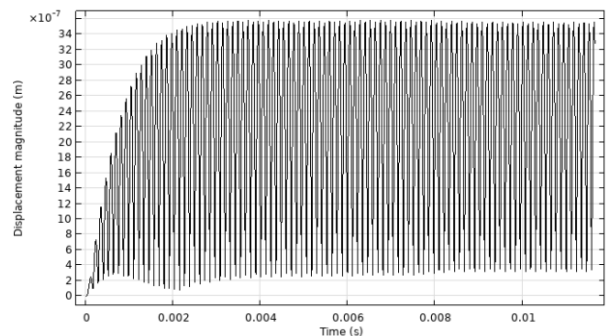

Fig. 12: Maximum Displacement for the Upper Surface (Damping Coefficient = 0.05)

It is observed that with the increase in the damping coefficient, the following changes occur:
1) The steady-state oscillation amplitude decreases.
2) The initial overshoot is reduced or eliminated.
3) The oscillation amplitude stabilizes more quickly.
All these observations logically align with the equations of vibration. In the subsequent phase, an equation is fitted to the motion profile of the piezo pump. This process utilizes software tools including SolidWorks, Get Data Graph Digitizer, and MATLAB. The coefficient 'a', which is time-dependent, is correlated with the maximum displacement of the piezoelectric surface.

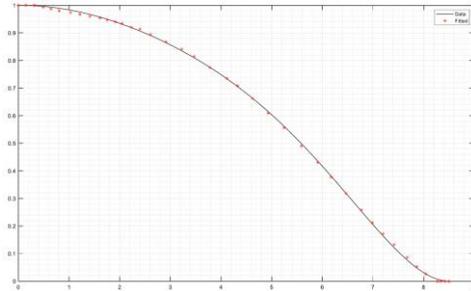

Fig. 13:The Fitted Curve (Distance Dependent Part)

In the above curve, the coefficient for maximum displacement is one. Therefore, the final equation of the curve is formulated, which depicts the position of each point on the curve as a function of time. The equation includes a time-dependent sinusoidal term with amplitude 'a' and a distance-dependent term represented by a sextic polynomial. This equation is particularly useful in CFD software for setting dynamic boundary conditions. Additionally, if velocity is required, one can derive it with respect to time from this equation.

To verify the consistency between the two-dimensional and three-dimensional solutions of the piezoelectric pump, simulations are conducted in four different scenarios: without damping, and with damping coefficients of 0.01, 0.03, and 0.05. Axisymmetric solutions are executed in each case, followed by a comparative analysis of the results. It is observed that the two-dimensional solution results closely approximate those of the three-dimensional solution. This approximation becomes more accurate with the increase in damping, enhancing the precision of the outcomes. Given the similarity of the results from both two-dimensional and three-dimensional solutions and the lower computational load of the two-dimensional approach, the latter is chosen for subsequent stages of the study.

**3.4 Micropump Geometry with Pump Room**
A careful examination of patents registered for piezoelectric pumps reveals that a critical component in these designs is the air chamber. Examples of these designs are illustrated in the figure below:

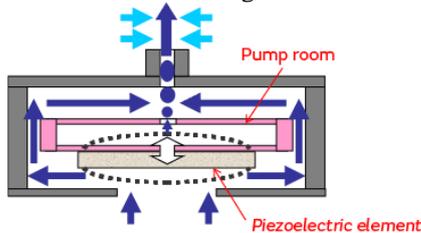

Fig. 14: Design of a Pump with a Pump Room

Upon revisiting the components of the reference pump, it is observed that a part assumed to function as an electrical conductor plays a significant role in controlling the pump's flow. The presence of this chamber enables the pump to efficiently transfer from a larger inlet area to a smaller outlet area, thereby minimizing the suction force required. The modeling of this component in SolidWorks software was carried out as follows:

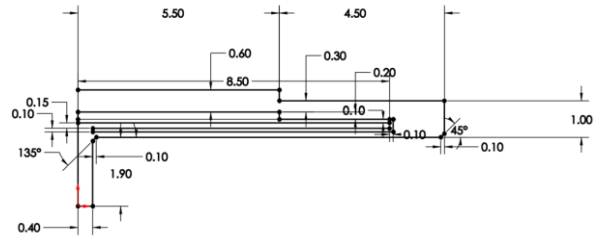

Fig. 15: 2D Modeling of Piezoelectric with Pump Room

**3.5 Integrated Simulation**
Due to computational cost and the adoption of one-way fluid-structure interaction (FSI) modeling, it was considered more feasible to use the fitted curve for displacement rather than employing structural and piezoelectric solvers. In this approach, the displacement of the structure is determined as a velocity in the Wall Movement section. The velocity of the structure is derived from the equation obtained in the previous sections. Due to the high speed of the fluid relative to the problem dimension, the Turbulent Flow, k-ε model was selected.

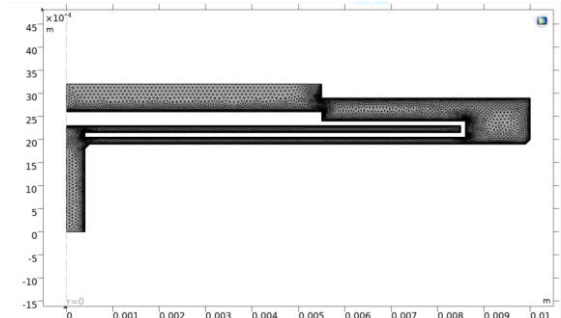

Fig. 16: New Geometry and Mesh (2D Axisymmetric)

**4. Results and Discussion**
In this section, the results of the comprehensive study conducted on the piezoelectric micropump are presented, emphasizing the key findings from the integrated multiphysics simulations and the operational efficiency of the pump has been demonstrated. These findings provide valuable guidelines for the optimization of design of piezoelectric micropumps.

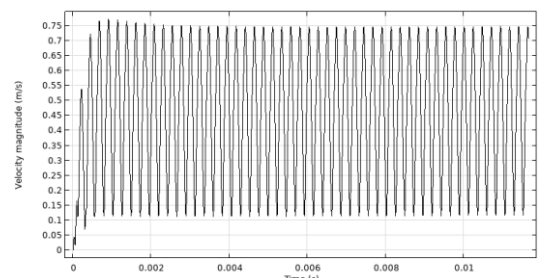

Fig. 17: Graph of Maximum Fluid Velocity on the Piezoelectric Surface

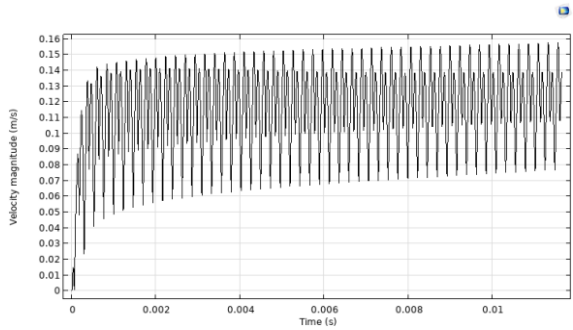
Fig. 18: Average Fluid Velocity at the Upper Boundary

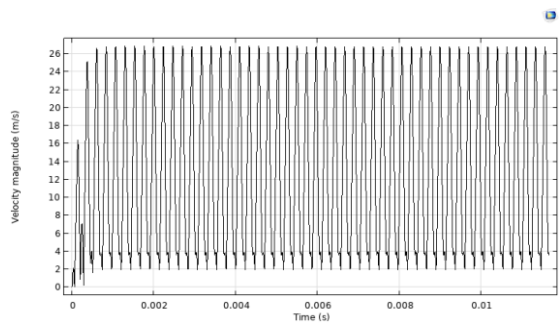
Fig. 19: Average Fluid Velocity at the Lower Boundary

Based on the above graphs, the pump's performance is observed to be accurate and highly efficient. Two selected frames from the animation are described as follows:

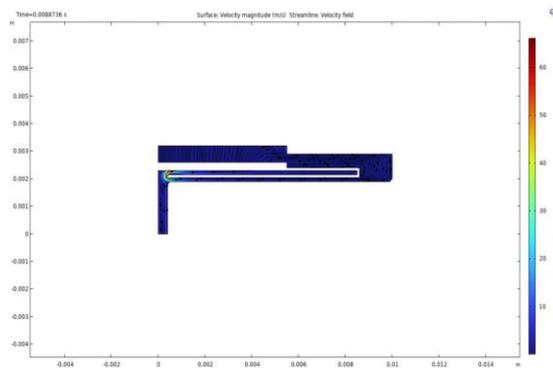
Fig. 20: Performance of the Pump in Suction Phase

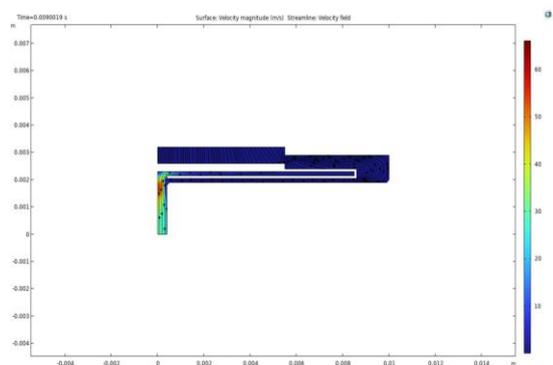
Fig. 21: Performance of the Pump in Pumping Phase

It is observed that the return flow has a significantly lower velocity compared to the outgoing flow. Additionally, the direction of the outgoing flow is perpendicular to the outlet, which is a desirable characteristic in the pump's design. This observation underscores the effectiveness of the pump's mechanism in controlling and directing the fluid flow, an essential aspect of its overall performance.

## 5. Concluding Remarks

The culmination of the study is manifested in the results, which provide a comprehensive understanding of the piezoelectric pump's performance characteristics. The integration of a new axisymmetric mesh and refined geometry, tailored to the specific requirements of the pump, allowed for a more accurate and efficient simulation process. The results highlight key aspects such as flow dynamics, displacement efficiency, and the overall effectiveness of the piezoelectric mechanism. These findings not only validate the theoretical models and design choices made but also offer valuable insights into potential areas for optimization. The successful implementation of the design strategies, including the introduction of a pump room and the adaptation to computational constraints, is clearly reflected in the performance metrics obtained. Overall, the results underscore the efficacy of the adopted methodologies and pave the way for future advancements in the design and application of piezoelectric pumps